\documentclass[aps,pra,onecolumn,groupedaddress,notitlepage,10pt]{revtex4-1}
\pdfoutput=1

\usepackage[T1]{fontenc}
\usepackage{graphicx}
\usepackage[labelfont=bf]{caption}
\usepackage{color}
\usepackage{hyperref}
\usepackage{amsmath}
\usepackage{amssymb}
\usepackage{setspace}

\newcommand{\ee}{\mathrm{e}}
\newcommand{\ii}{\mathrm{i}}
\renewcommand{\eqref}[1]{Eq.~(\ref{#1})}

\begin{document}


\title{A plane wave analysis of coherent holographic image reconstruction by phase transfer}


\author{Jeffrey J.~Field}
\affiliation{W.M. Keck Laboratory for Raman Imaging of Cell-to-Cell Communications, Colorado State University, Fort Collins, CO 80523}
\affiliation{Department of Electrical and Computer Engineering, Colorado State University, Fort Collins, CO 80523 \\ \href{mailto:jeff.field@colostate.edu}{\color{blue} jeff.field@colostate.edu}}

\author{David G.~Winters}
\affiliation{W.M. Keck Laboratory for Raman Imaging of Cell-to-Cell Communications, Colorado State University, Fort Collins, CO 80523}
\affiliation{Department of Electrical and Computer Engineering, Colorado State University, Fort Collins, CO 80523}

\author{Randy A.~Bartels}
\affiliation{W.M. Keck Laboratory for Raman Imaging of Cell-to-Cell Communications, Colorado State University, Fort Collins, CO 80523}
\affiliation{Department of Electrical and Computer Engineering, Colorado State University, Fort Collins, CO 80523}
\affiliation{School of Biomedical Engineering, Colorado State University, Fort Collins, CO 80523}


\date{\today}

\begin{abstract}
Fluorescent imaging plays a critical role in a myriad of scientific endeavors, particularly in the biological sciences. Three-dimensional imaging of fluorescent intensity often requires serial data acquisition, that is voxel-by-voxel collection of fluorescent light emitted throughout the specimen with a non-imaging single-element detector. While non-imaging fluorescence detection offers some measure of scattering robustness, the rate at which dynamic specimens can be imaged is severely limited. Other fluorescent imaging techniques utilize imaging detection to enhance collection rates. A notable example is light-sheet fluorescence microscopy, also known as selective-plane illumination microscopy (SPIM), which illuminates a large region within the specimen and collects emitted fluorescent light at an angle either perpendicular or oblique to the illumination light sheet. Unfortunately, scattering of the emitted fluorescent light can cause blurring of the collected images in highly turbid biological media. We recently introduced an imaging technique called coherent holographic image reconstruction by phase transfer (CHIRPT) that combines light-sheet-like illumination with non-imaging fluorescent light detection. By combining the speed of light-sheet illumination with the scattering robustness of non-imaging detection, CHIRPT is poised to have a dramatic impact on biological imaging, particularly for {\em in vivo} preparations. Here we present the mathematical formalism for CHIRPT imaging under spatially coherent illumination and present experimental data that verifies the theoretical model.
\end{abstract}


\maketitle

\section{Background}

Image formation by temporal modulation of illumination light has emerged as a promising method to circumvent current limitations in biological imaging applications \cite{Sanders:1991as,Feldkhun:2010vm,Futia:2011tra,Schlup:2011gpa,Howard:2012gca,Higley:2012ce,Higley:2013vfa,Diebold:2013uy,Hwang:2015xe,Winters:2015rt,Field:2015fr}.  In particular, image acquisition speeds can be greatly enhanced by overcoming the need to serially collect three-dimensional image data, as is required for techniques such as confocal and multiphoton laser scanning microscopy (MPLSM) \cite{Davidovits:1969td,Denk:1990bu}. Imaging systems relying on temporal modulations of the illuminating light to form images collect image data from multiple points in the specimen simultaneously with a single-element detector. 

Recently, we introduced an imaging method we call CHIRPT, for coherent holographic image reconstruction by phase transfer, which allows coherent imaging methods, including holographic propagation, to be directly applied to incoherent contrast mechanisms such as fluorescence \cite{Field:2015fr}.  CHIRPT has demonstrated an ability to dramatically increase imaging speeds and the depth of field when compared to conventional biological imaging methods.  

In this Paper, we rigorously investigate the process of image formation in CHIRPT by an angular spectrum representation \cite{Novotny:2010qd} to compute to the illumination intensity pattern.  We then compute the transfer function and point spread function (PSF) of an idealized CHIRPT imaging system, and show how this allows the spatial phase of the imaging system to be recovered with a non-iterative algorithm.  Finally, we verify the theoretical model with an experimentally determined transfer function.

\section{Principles of CHIRPT Imaging}
CHIRPT forms images by illuminating a large area in the object region with a temporally modulated intensity pattern that is unique to each point in the lateral-axial plane, $(x,z)$.  The contrast mechanism, which can be coherent or incoherent, responds to this modulation pattern and re-emits, transmits, or scatters signal light that has the same spatio-temporal structure as the illumination intensity.  The spatio-temporally modulated signal light is measured with a single-element photodetector, such as a photomultiplier tube (PMT), providing a periodic voltage signal, $S(t)$.  The electronic spectrum of this signal, $\tilde{S}(\nu)$, where $\nu$ is the temporal modulation frequency, encodes a one-dimensional image of the object while preserving the spatial phase.  A two-dimensional image in the $(x,z)$ plane is reconstructed by coherently propagating the contrast intensity, which relies on the spatial phase of the illumination fields preserved in $S(t)$.

\begin{figure*}[ht]
\begin{center}
\resizebox{\linewidth}{!}{\includegraphics{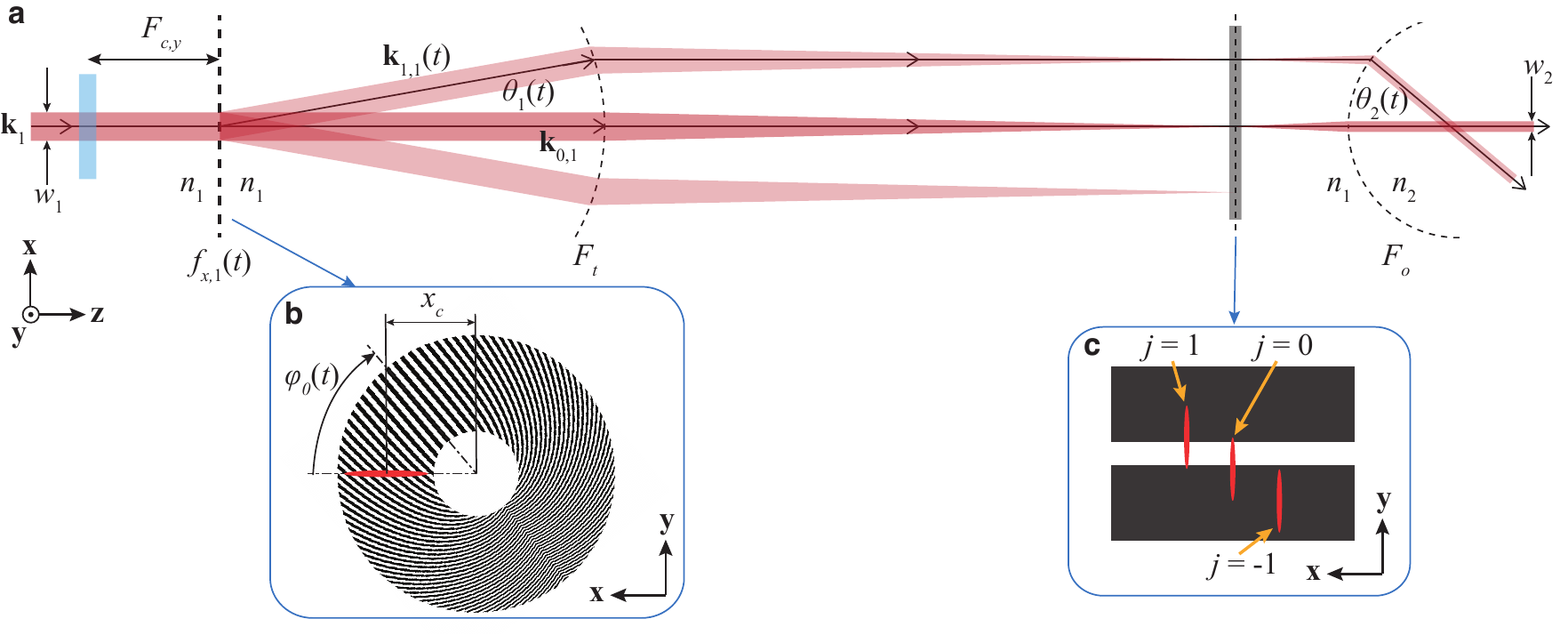}}
\caption{\label{fig:schematic} Schematic representation of a CHIRPT microscope. (a) Image relay system in the $(x,z)$ plane. (b) A modulation mask of the form $m(R,\varphi) = 1/2 + (1/2) \, \mathrm{sgn}\left[ \cos (\Delta k \, R \, \varphi ) \right]$ with $\Delta k = 3/\mathrm{mm}$. (c) To provide a unique mapping of each $(x,z)$ location in the object region, only the undiffracted beam and one of the first-order diffracted beams are permitted to propagate to the object plane. A horizontally-oriented slit is placed in the pupil plane to completely block one of the diffracted beams.}
\end{center}
\end{figure*}

A CHIRPT microscope is represented schematically in Fig.~\ref{fig:schematic}. A collimated, spatially-coherent illumination beam is brought to a line focus on the face of a circular modulation mask using a cylindrical lens of focal length $F_{c,y}$, oriented to focus the beam in the vertical dimension (${\bf y}$).  The mask is designed such that the modulation frequency imparted to the line focus varies linearly with a function of radial position, $R$ (Fig.~\ref{fig:schematic}b). The varying spatial frequency presented to the line focus, $f_{x,1}(t)$, varies as the mask is rotated at a constant angular velocity, causing the illumination beam to be diffracted from the mask with an angle that varies with time, $\theta_1(t) = \sin^{-1} \left[ \lambda \, f_{x,1}(t) / n_1 \right]$. An imaging system is used to image relay the modulation plane to the specimen.  Schematically, we represent the imaging system as a tube lens of focal length $F_t$ and an objective lens of focal length $F_o$, providing a system magnification of $M = F_t/F_o$. 

The critical component of a CHIRPT imaging system is the illumination intensity pattern, $I_\mathrm{ill}({\bf r},t)$, which illuminates every position in the $(x,z)$ plane in the specimen with a unique intensity modulation as a function the rotation angle of the mask, $\varphi_0(t)$. For the illumination intensity in the object region to meet this criteria, only the undiffracted beam and one of the first-order diffracted beams are permitted to interfere in the specimen region. This is accomplished with a horizontally-oriented slit placed in the conjugate image plane, which is possible due to a constant non-zero spatial frequency imparted to the illumination beam in the vertical dimension, $y$ (Fig.~\ref{fig:schematic}c).  This spatial frequency is computed in Section~\ref{sec:spatial_freq}.  

As the mask completes a full rotation, the full numerical aperture of the imaging system is scanned by the diffracted beam. Each temporal measurement on the photodetector corresponds to a specific angle of the diffracted beam in the object region, $\theta_2(t)$. The diffracted beam interferes with the undiffracted beam and creates an interference patten with a single spatial frequency in the lateral dimension, $f_{x,2}(t)$. This produces the temporal photodetector signal as a function of time corresponding to a measurement of the lateral spatial frequency content of the specimen, and a one-dimensional image in the lateral dimension can be recovered by a Fourier transform. As we show in the following analysis, spatial phase information is also encoded into this signal, and a two-dimensional image in the $(x,z)$ plane can be recovered from a temporal measurement that corresponds to a single rotation of the modulation mask.

\section{Theoretical Analysis}

CHIRPT encodes complex spatial image information into a temporal intensity by measuring the radiant flux from the contrast distribution on a single-element photodetector as the illumination intensity pattern is modulated over time.  The CHIRPT signal has the form of a projection:
\begin{equation}
S(t) = \int_{-\infty}^{\infty} \! \! \mathrm{d}^3 {\bf r} \, \, I_\mathrm{ill}({\bf r},t) \, C({\bf r}, t) = \left< I_\mathrm{ill}({\bf r},t) \, C({\bf r}, t) \right>_{\bf r}
\end{equation}
where $I_\mathrm{ill}({\bf r},t)$ is the illumination pattern in the object region, and $C({\bf r}, t)$ is the contrast distribution.  The contrast distribution describes the method by which illumination intensity is transferred to contrast intensity.  For example, the contrast distribution can simply represent the transmittance of the object if the measured contrast intensity is the illumination source after propagation through an object displaying absorption.  In that case, $C({\bf r},t) = \mathbb{T}({\bf r},t)=1-\mathbb{A}({\bf r},t) $, where $\mathbb{T}$ is the intensity transmission of the object, and $\mathbb{A}$ is the intensity absorption of the object, and an image is formed from the object transmission.  The contrast can also represent the spatial concentration distribution of fluorophores in an object, scaled by the absorption cross section for the excitation light and efficiency of fluorescent emission.  Throughout this analysis, we shall assume that the contrast function is constant for the duration of a scan, so that $C({\bf r}, t) \rightarrow C({\bf r})$.  We also assume that the contrast function is linearly proportional to the illumination intensity.

The propagation phase imparted to the intensity emerging from the object arises from a change in the instantaneous modulation frequency at a given point in the contrast distribution as a function of scan time.  In general, the spatial phase of the illumination source is encoded in the signal from the photodetector as a phase variation on the temporal carrier frequency.

This phase transfer is a critical aspect of CHIRPT. Here, we develop a theory that describes the generation of CHIRPT signals. To derive the form of the signal generated by the photodetector, we consider a spatially-coherent line focus passing through a spatio-temporally varying modulation mask.  Diffraction by the mask produces transverse wave vectors that vary with time.  We model this diffraction by a set of plane waves propagating at angles with respect to the optic axis ($z$) that vary with time (Fig.~\ref{fig:schematic}).  Using these angles, an expression for the electric field diffracted from the modulation mask for an arbitrary set of diffracted fields is derived.  The phase accumulated upon propagation into the object region is computed via angular spectrum formalism applied to the re-imaged diffracted waves that interfere to produce the illumination intensity, taking the system magnification into account.  An expression for the total illumination field in the object region is found by taking the square-magnitude of the sum of the diffracted orders. A  pupil phase is included to account for systematic aberrations to the imaging system.

\subsection{Plane Wave Expansion of the Illumination Intensity Pattern}
CHIRPT is part of a broader class of imaging in which multiple diffractive orders are used to form a time-dependent illumination intensity pattern which encodes image information into a whole-field measurement.  Other methods that fall in this category include synthetic aperture microscopy (SAM) \cite{Alexandrov:2006iy,Ralston:2007hsa,Davis:2008dy,Hillman:2009wg,Feldkhun:2010vm} and spatial-frequency-modulation for imaging (SPIFI) \cite{Schlup:2011gpa,Futia:2011tra,Higley:2012ce,Howard:2012gca,Higley:2013vfa}.  As such, we begin by deriving general expressions for the illumination intensity in the object region when a spatially coherent illumination source is modulated by an arbitrary mask pattern.

In the following analysis, we use the subscript pair $(j,s)$ to denote the $j^{\mathrm{th}}$ diffracted order and either the mask region with $s = 1$ or the object region with $s = 2$. For example, the set of k-vectors corresponding to the diffracted beams in the mask region are described by wavevectors ${\bf k}_{j,1}  =  k_{x,j,1} \, \hat{\bf e}_x + k_{y,j,1} \, \hat{\bf e}_y + k_{z,j,1}  \, \hat{\bf e}_z$. Spatial coordinates are described with an arbitrary position vector, ${\bf r} = x \, {\hat{\bf e}}_x +  y \, {\hat{\bf e}}_y +  z \, {\hat{\bf e}}_z$. The optic axis of the system is set to be the z-direction ($\hat{\bf e}_z$), and we define transverse k-vectors (in the $x , y$ plane) by the notation ${\bf k}_{\perp,j,s} = k_{x,j,s} ~\hat{\bf e}_x + k_{y,j,s} ~\hat{\bf e}_y$ and the position in the transverse plane as ${\bf r}_{\perp} = x ~ \hat{\bf e}_x +  y ~ \hat{\bf e}_y$.

For simplicity, we assume a uniform plane wave of unit amplitude incident on a diffraction mask, and propagating along the optic axis, which is normal to the plane that defines the diffraction mask. The incident field contains a single spatial frequency, so the spatial portion of the electric field incident on the modulation mask can be written as:
\begin{equation}
E_\mathrm{inc}({\bf r}) = \mathrm{e}^{\mathrm{i} \, {\bf k}_1 \cdot {\bf r}} =  \mathrm{e}^{\mathrm{i} \, k_1 \, z}
\end{equation}
where $k_1 = 2 \pi \, n_1 / \lambda$ is the wavenumber of the incident wave and the wave vector is strictly along z: ${\bf k}_1 =   k_1 \, \hat{\bf{e}}_z$. 

The modulation mask causes the incident field to diffract into multiple directions propagating at various angles with respect to the optic axis.  The structure of the mask dictates the direction of the diffracted wave vectors by imparting a transverse wavevector to the incident plane wave.  Expressing the magnitude of the k-vector in the transverse plane $(x,y)$ of a diffracted order $j$ as $\left| {\bf k}_{\perp,j,s}(t) \right| = k_{\perp,j,s}(t)$, the magnitude of the axial k-vector for each diffracted wave in the mask region is:
\begin{equation}
k_{z,j,1}(t) = \left| {\bf k}_{z,j,1}(t) \right| = \sqrt{k_1^2 - {\bf k}_{\perp,j,1}(t)\cdot {\bf k}_{\perp,j,1}^*(t)} = k_1 \, \sqrt{ 1 -  \left[ \frac{k_{\perp,j,1}(t)}{k_1} \right]^2 }.
\label{k_ewald}
\end{equation}
We can therefore write the electric field for the $j^\mathrm{th}$ diffracted wave in the mask region as:
\begin{equation}
v_{j,1}({\bf r},t) = a_j \, \mathrm{e}^{\mathrm{i} \, {\bf k}_{j,1}(t) \cdot {\bf r} }  = a_j \, \exp \left[ \mathrm{i} \, {\bf k}_{\perp,j,1}(t) \cdot {\bf r}_{\perp} \right]  \, \exp \left[\mathrm{i} \, k_{z,j,1}(t) \, z \right].
\end{equation}
where the amplitude of the diffracted field can be a complex quantity, i.e., $a_j = |a_j| \mathrm{e}^{\mathrm{i} \, \angle a_j}$. Physically, the phase of the amplitude, $\angle a_j$, corresponds to a lateral translation of the modulation mask. 

The electric field distribution directly behind the modulator is image relayed to the object plane with an imaging system of magnification $M$ (Fig.~\ref{fig:schematic}). Therefore the axial wavevector in the object plane is:
\begin{equation}
k_{z,j,2}(t) = \sqrt{k_{2}^2 - \left[ k_{\perp,j,2}(t) \right]^2} = k_2 \, \sqrt{1 - \left[\frac{n_2}{n_1} \, \frac{M \, k_{\perp,j,1}(t)}{k_2} \right]^2} 
\end{equation}
where we have used the fact that $k_{\perp,j,2}(t) = (n_2/n_1) \, M \, k_{\perp,j,1}(t)$, and the wavenumber in the object region is scaled by the index of the immersion medium, $n_2$.  Thus the electric field for the $j^\mathrm{th}$ diffracted wave in the object plane becomes:
\begin{equation}
v_{j,2}({\bf r},t) =  a_{j} \, \exp \left[\mathrm{i} \, \frac{n_2}{n_1} \, M \, {\bf k}_{\perp,j,1}(t) \cdot {\bf r}_\perp \right]  \,  \exp \left\{ \mathrm{i} \, k_2 \, z \, \sqrt{1 - \left[\frac{n_2}{n_1} \, \frac{M \, k_{\perp,j,1}(t)}{k_2} \right]^2} \right\} 
\label{general_field}
\end{equation}
While \eqref{general_field} accounts for the phase caused by diffraction away from the focal plane (system defocus), there is, in general, an arbitrary pupil phase, $\Phi_\mathrm{pupil}\left[{\bf k}_{\perp,j}(t)\right]$, that must be included to account for other spatial phase variations imparted by the imaging system.  Including the pupil phase, the expression for a diffracted order in the object region is:
\begin{equation}
v_{j,2}({\bf r},t) = a_{j} \, \exp \left[\mathrm{i} \, {\bf k}_{\perp,j,2}(t) \cdot {\bf r}_\perp \right] \, \exp \left[ \mathrm{i} \, k_{z,j,2}(t) \, z \right]  \, \exp \left\{ \mathrm{i} \, \Phi_\mathrm{pupil}\left[{\bf k}_{\perp,j}(t)\right]\right\} 
\end{equation}

In the object region, the diffracted beams ($j = \pm1$) propagate at angles $\pm \theta_2(t) = \sin^{-1}\left[ \pm \lambda \, k_{\perp,j,2}(t) / (2 \pi) \right] $, so that we need to account for diffracted waves at $\pm {\bf k}_{\perp, j,2}(t)$. A given mask can generate many diffracted waves, which can either be derived from higher diffracted orders of a particular mask lattice vector, or from the inclusion of multiple lattice vectors that can diffract beams to additional angles \cite{Betzig:2005jk}. If we suppose that we have a total of $N$ diffracted waves of appreciable power, then we have a total of $2N +1$ waves (including the zero-order term) that must be accounted for. The total illumination field propagating in the object region can thus be compactly expressed as
\begin{equation}
E_\mathrm{ill}({\bf r},t)  =  \sum_{j=-N}^{N} v_{j,2}({\bf r},t) 
\label{eq:efield_compact}
\end{equation}

Since the mask is real and positive semidefinite, all of the diffracted waves for given index $j$ have conjugate amplitude, i.e., $-a_{j}=a_{j}^*$. Also, each pair of diffracted waves $\pm j$ have the same axial wave vector, but conjugate transverse wavevectors, so conjugate pairs in the summation can be paired:
\begin{eqnarray}
E_\mathrm{ill}({\bf r},t) & = & a_0 \, \mathrm{e}^{\mathrm{i} \, k_2 \, z}\, \mathrm{e}^{\mathrm{i} \, \Phi_\mathrm{pupil}(0)} \nonumber \\
	& & + 2 \sum_{j=1}^{N}  \left| a_j \right| \cos \! \left[\frac{n_2}{n_1} \, M \, {\bf k}_{\perp,j,1}(t) \cdot {\bf r}_\perp + \angle{a_j} \right]  \exp \! \left\{ \mathrm{i} \, k_2 \, z \, \sqrt{1 - \left[\frac{n_2}{n_1} \, \frac{M \, k_{\perp,j,1}(t)}{k_2} \right]^2} \right\} \,  \mathrm{e}^{\mathrm{i} \, \Phi_\mathrm{pupil}\left[{\bf k}_{\perp,j}(t) \right] }
\label{intensity-spifi}
\end{eqnarray}
Here the $j=0$ term is the undiffracted beam. The illumination intensity in the object region is the square modulus of the expression above, $I_\mathrm{ill}({\bf r},t) = \left| E_\mathrm{ill}({\bf r},t) \right|^2$. Defining the complex conjugate of the $j^\mathrm{th}$ electric field as:
\begin{equation}
u_{j,2}({\bf r},t) \equiv \left[ v_{j,2}({\bf r},t) \right]^{*}
\end{equation}
the illumination intensity can be expressed as:
\begin{equation}
I_\mathrm{ill}({\bf r},t) = \sum \limits_{j = -N}^{N} \sum \limits_{k = -N}^{N} v_{j,2}({\bf r},t) \, u_{k,2}({\bf r},t)
\label{eq:int_sum}
\end{equation}
where $N$ is the highest diffracted order included in the intensity pattern.

\subsection{Plane Wave Expansion of the Modulation Mask \label{sec:spatial_freq}}
Utilizing \eqref{intensity-spifi}, the illumination intensity in the object region for an arbitrary modulation mask can be found by computing the set of transverse wavevectors ${\bf k}_{\perp,j,1}(t)$ imparted on the illumination beam by the mask.  To obtain an expression for the transverse wavevectors, we consider the mask as a decomposition of plane wave vectors.  We assume that the mask is constructed of periodic modulations along directions described by lattice vectors $\hat{{\bf e}}_p$. We further assume that the mask modulates the amplitude of the field, requiring that it be positive semidefinite.  With both of these assumptions in hand we can write the general form of the mask as a Fourier cosine decomposition of wavevectors:
\begin{equation}
m({\bf r},t) = b_0 + \sum_{p = 1}^{\infty} \, b_p \, \cos \left[{\bf k}_p(t) \cdot {\bf r} + \alpha_p(t) \right]
\end{equation}
where we have utilized the fact that ${\bf k}_p(t) = k_p(t) \, \hat{{\bf e}}_p$, and $\alpha_p(t)$ is a phase shift that accounts for lateral translations of the modulation mask as a function of scan time. The term $b_0$ is simply a constant offset that must be selected to allow the mask to be positive semidefinite.  All other terms correspond to periodic patterns in the modulation mask. 

Since the modulation mask is confined to the lateral plane, i.e., the mask provides no axial modulations, this wavevector is equivalent to the transverse wavevector experienced by the incident electric field.  Therefore, ${\bf k}_p(t) = {\bf k}_{\perp,j,1}(t)$, and we arrive at a general expression for the modulation mask in terms of the lateral wavevectors imparted onto the incident coherent illumination source:
\begin{equation}
m({\bf r}_\perp,t) = b_0 + \sum_{j=1}^{N} \, b_j \, \cos \! \left[{\bf k}_{\perp,j,1}(t) \cdot {\bf r}_\perp + \alpha_j(t) \right]
\label{eq:mask_general}
\end{equation}
For $j \neq 0$, $b_j = 2 |a_j|$.

\section{CHIRPT Imaging \label{sec:chirpt_imaging}}
We now consider the form of the modulation mask used for CHIRPT imaging.  In polar coordinates, $(R,\varphi)$, the pattern of the CHIRPT modulation mask has the form:
\begin{equation}
m(R,\varphi)= \frac{1}{2} + \frac{1}{2} \mathrm{sgn} \! \left[ \cos \! \left(\Delta k \, R \, \varphi \right) \right]
\label{basic-mask}
\end{equation}
where $\mathrm{sgn} \! \left[ \cdot \right]$ is included to account for the binary amplitude modulation resulting from the printing process, and $\Delta k$ sets the highest density of the printed pattern. An example mask is shown in Fig.~\ref{fig:schematic}b.

We wish to decompose the transverse wave vector imparted on the cylindrically-focused beam into Cartesian components, where $x$ is the lateral dimension, parallel to the line focus, and $y$ is the vertical direction, perpendicular to the line focus.  Equation~\eqref{basic-mask} represents the full two-dimensional pattern of the mask, while we are interested in the region sampled by the line focus, located at $y=0$, and approximately centered on one side of the mask such that the illumination distribution in $x$ lies between the inner and outer radii of the mask (Fig.~\ref{fig:schematic}b).  To account for the changing modulation pattern sampled by the line focus during mask rotation, and thus find the local spatial frequencies as a function of scan time, we consider the mask:
\begin{equation}
m\left[R,\varphi - \varphi_0(t)\right] = \frac{1}{2} + \frac{1}{2} \mathrm{sgn} \! \left( \cos \! \left\{ \Delta k \, R \, \left[\varphi - \varphi_0(t) \right] \right\} \right) \end{equation}
where $\varphi_0(t)$ describes the rotation angle of the mask.  Since the mask rotates at a constant angular velocity, $\varphi_0(t) = 2 \pi \, \nu_r \, t$, where $\nu_r$ is the rotational frequency of the mask.

To uniquely determine the spatial phase of the illumination microscope, only the first diffracted order from the mask is permitted to interfere with the undiffracted beam in the object plane.  Higher diffracted orders resulting from the binary modulation scheme are omitted in the following analysis, and also excluded in our CHIRPT experimental design, so that the mask we consider is:
\begin{eqnarray}
m\left[R,\varphi - \varphi_0(t)\right] & = & \frac{1}{2} + \frac{1}{2} \cos \! \left\{ \Delta k \, R \, \left[\varphi - \varphi_0(t) \right] \right\} \nonumber \\
	& = & \frac{1}{2} + \frac{1}{2} \cos \! \left[\, \phi \! \left( R, \varphi, t \right) \, \right]
\end{eqnarray}

The local linear spatial frequency imparted by the mask in Cartesian coordinates is \cite{Goodman:2005jt}:
\begin{eqnarray}
f_{x,1}(t) & = & \frac{1}{2 \pi} \, \frac{\partial \phi ({\bf r}, \varphi, t)}{ \partial x}\\
f_{y,1}(t) & = & \frac{1}{2 \pi} \, \frac{\partial \phi ({\bf r}, \varphi, t)}{ \partial y}
\end{eqnarray}
The region sampled by the beam is along the line where $\varphi = 0$, so the local angular spatial frequencies can be calculated as:
\begin{eqnarray}
k_{x,1}(t) & = &  \left[ \cos \! \varphi \, \frac{\partial \phi(R, \varphi, t)}{\partial r}  - \frac{1}{r} \, \sin \! \varphi \, \frac{\partial \phi(R, \varphi, t)}{\partial \, \varphi}  \right]_{\varphi = 0}  \\
k_{y,1}(t) & = & \left[ \sin \! \varphi \, \frac{\partial \phi(R, \varphi, t)}{\partial r}  + \frac{1}{r} \, \cos \! \varphi \, \frac{\partial \phi(R, \varphi, t)}{\partial \, \varphi}  \right]_{\varphi = 0}
\end{eqnarray}
where we have converted linear spatial frequency to angular spatial frequency by multiplying by $2\pi$.  The local spatial frequencies evaluate to:
\begin{eqnarray}
k_{x,1}(t) & = & 2 \pi \,\Delta k \, \nu_r \, t \equiv 2 \pi \, \kappa_1 \, t \label{lateral_spatial_freq} \\
k_{y,1}(t) & = & \Delta k \label{vertical_spatial_freq}
\end{eqnarray}
and the transverse wave vector for the fundamental mask frequency is ${\bf k}_{\perp,1,1}(t) = 2 \pi \, \kappa_1 \, t \, \hat{{\bf e}}_x + \Delta k \, \hat{{\bf e}}_y$. The parameter $\kappa_1$ is the so-called chirp parameter, which relates the modulation frequency to lateral position on the modulation mask.

The expressions above elucidate the key elements of CHIRPT imaging.  In the lateral dimension, spatio-temporal modulations encode the lateral position with unique modulation frequencies.  Moreover, each measurement in time corresponds to a unique lateral spatial frequency, demonstrating how CHIRPT measures the lateral spatial frequency content of the object.

Returning to the general form of the modulation mask in \eqref{eq:mask_general}, the transverse wavevector can be decomposed into Cartesian components to give:
\begin{equation}
m(x,y,t) =  b_0 +  \sum_{j=1}^{N} b_j \, \cos \! \left[j \, 2 \pi \, \kappa_1 \, t \, x + j \, \Delta k \, y + \alpha_j(t) \right]
\end{equation}
Since CHIRPT only utilizes one of the first-order diffracted modes and the undiffracted beam, the plane wave expansion of the mask can be simplified further:
\begin{equation}
m(x,y,t) =  b_0 + b_1 \cos \! \left[2 \pi \, \kappa_1 \, t \, x + \Delta k \, y + \alpha_1(t) \right]
\end{equation}
Since the mask contains a finite set of spatial frequencies, the function describing the mask must be constrained to a finite time window. Thus we can write the mask as:
\begin{equation}
m(x,y,t) = w(t) \, \left\{ b_0 + b_1 \cos \! \left[2 \pi \, \kappa_1 \, t \, x + \Delta k \, y + \alpha_1(t) \right] \right\}
\end{equation}
where the temporal window function $w(t)$ restricts the mask function to a single rotation of the disk and is used to describe apodization of the throughput energy. Throughout this analysis we assume that $w(t)$ is a real-valued function. 

As noted previously, the CHIRPT illumination intensity pattern is formed by allowing only one of the diffracted orders to interfere with the undiffracted beam in the object region.  This is accomplished with a horizontally-oriented slit placed near the pupil plane of the illumination lens (Fig.~\ref{fig:schematic}c) that passes only a portion of the $j=0$ and $j=1$ beams, which is possible due to a non-zero spatial frequency component in the vertical ($\hat{\bf e}_y$) direction that is independent of the rotation angle of the mask [\eqref{vertical_spatial_freq}].  Including only the $j = +1$ and $j = 0$ orders the illumination field [\eqref{eq:efield_compact}] is $E_{\mathrm{ill}}({\bf r},t) = v_{0,2}({\bf r},t) + v_{1,2}({\bf r},t)$, where:
\begin{eqnarray}
v_{0,2}({\bf r},t) & = & a_0 \,  \exp \left[\mathrm{i} \, k_2 \, z \right] \, \exp \left[\mathrm{i} \, \Phi_\mathrm{pupil}(0) \right] \\
v_{1,2}({\bf r},t) & = & a_1 \, w(t) \, \exp \left( \mathrm{i} \, 2 \pi \, \frac{n_2}{n_1} \, M \, \kappa_1 \, t \, x \right) \, \exp \left( \mathrm{i} \, \frac{n_2}{n_1} \, M \, \Delta k \, y \right)  \nonumber \\  
	& & \times \exp \! \left[\mathrm{i} \, k_2 \, z \, \sqrt{1 - \left(\frac{\lambda}{n_1} \, M \, \kappa_1 \, t \right)^2 - \left(\frac{n_2}{n_1} \, \frac{M \, \Delta k}{k_2} \right)^2} \right]  \nonumber \\
	& & \times \exp \left[\mathrm{i} \, \Phi_\mathrm{pupil}(t) \right] \, \exp \left[\mathrm{i} \, \alpha_1(t) \right]
	\label{eq:first_order_field}
\end{eqnarray}
Note that $w(t)$ does not apply to $v_{0,2}({\bf r},t)$ because this field is stationary with scan time. 

As indicated in Fig.~\ref{fig:schematic}b, the optic axis of the illumination beam is not collinear with the center of the mask modulation pattern. The consequence is that the illumination intensity, and hence the CHIRPT signal, contains a non-zero carrier frequency. This is a critical aspect of CHIRPT imaging, since a non-zero carrier frequency permits separation of the positive and negative sidebands in modulation frequency space, allowing single-shot data acquisition. Following the notation outlined in Fig.~\ref{fig:schematic}b, we let the lateral position in Eq.~\eqref{eq:first_order_field} be $x = x_c + \Delta x$, where $x_c$ denotes the location of the centroid of the illumination beam intensity in the lateral dimension in the mask plane. Defining the carrier frequency as $\nu_c = (n_2/n_1) \, M \, \kappa_1 \, x_c$, we can rewrite Eq.~\eqref{eq:first_order_field} as:
\begin{eqnarray}
v_{1,2}({\bf r},t) & = & a_1 \, w(t) \, \exp \, \left( \mathrm{i} \, 2 \pi \, \nu_c \, t \right) \exp \left( \mathrm{i} \, 2 \pi \, \frac{n_2}{n_1} \, M \, \kappa_1 \, t \, \Delta x \right) \, \exp \left( \mathrm{i} \, \frac{n_2}{n_1} \, M \, \Delta k \, y \right)  \nonumber \\  
	& & \times \exp \! \left[\mathrm{i} \, k_2 \, z \, \sqrt{1 - \left(\frac{\lambda}{n_1} \, M \, \kappa_1 \, t \right)^2 - \left(\frac{n_2}{n_1} \,  \frac{M \, \Delta k}{k_2} \right)^2} \right]  \nonumber \\
	& & \times \exp \left[\mathrm{i} \, \Phi_\mathrm{pupil}(t) \right] \, \exp \left[\mathrm{i} \, \alpha_1(t) \right]
\end{eqnarray}

The illumination intensity is defined as the square modulus of the illumination field, $I_\mathrm{ill}({\bf r},t) = \left| E_{\mathrm{ill}}({\bf r},t)  \right|^2$. 
We may therefore write the illumination intensity in the object region as:
\begin{equation}
I_\mathrm{ill}({\bf r},t) = v_{0,2}({\bf r},t) u_{0,2}({\bf r},t) + v_{1,2}({\bf r},t) u_{1,2}({\bf r},t) + v_{0,2}({\bf r},t) u_{1,2}({\bf r},t) + u_{0,2}({\bf r},t) v_{1,2}({\bf r},t)  
\label{eq:foil_fields}
\end{equation}
For brevity, we write the electric field for each beam as:
\begin{eqnarray}
v_{0,2}({\bf r},t) & = & a_0 \, \ee^{\ii \, \Phi_{0,2}({\bf r},t)} \\
v_{1,2}({\bf r},t) & = & a_1 \, w(t) \, \ee^{\ii \, \Phi_{1,2}({\bf r},t)} 
\end{eqnarray}
where the overall phase argument of diffracted beam $j$ is $\Phi_{j,2}({\bf r},t)$. The illumination intensity becomes:
\begin{eqnarray}
I_\mathrm{ill}({\bf r},t) & = & a_0^2 + a_1^2  w^2(t) + 2 \, a_0 \, a_1 \,  w(t) \, \cos \left[\Phi_{1,2}({\bf r},t) - \Phi_{0,2}({\bf r},t) \right] \nonumber \\
	& = & a_0^2 + a_1^2  w^2(t) + 2 \, a_0 \, a_1 \,  w(t)  \, \cos \left[ \Delta \Phi ({\bf r},t) \right]
\label{eq:intensity_generic}
\end{eqnarray}

Equation~\eqref{eq:intensity_generic} elucidates the key principle of CHIRPT imaging -- namely, that the illumination intensity, and hence the CHIRPT signal, is formed by the sum to two separate intensity components: one with no dependence on the phase of the illumination beams, and therefore no carrier frequency, and another with spatio-temporal intensity fringes that depend on the phase difference between the two illumination beams, $\Delta \Phi ({\bf r},t)$. Since the dependence of the CHIRPT signal on the phase difference of the illumination beams is independent of the contrast mechanism, coherent imaging methods can be applied directly to incoherent light for the first time.

The overall phase of each beam results from the varied optical path length (OPL) that each beam experiences while propagating from the modulation mask to a position ${\bf r}$ in the object region. The OPL and the phase are related by the wavelength of the illumination light by:
\begin{equation}
\Phi_{j,2}({\bf r},t) = \frac{2 \pi}{\lambda} \, \mathrm{OPL}_{j,2}({\bf r},t)
\end{equation}
Clearly the intensity modulation pattern is driven by the difference in OPL experienced by the two illumination beams:
\begin{equation}
\Delta \Phi({\bf r},t) = \frac{2 \pi}{\lambda} \left[ \mathrm{OPL}_{1,2}({\bf r},t) - \mathrm{OPL}_{0,2}({\bf r},t) \right] = \frac{2 \pi}{\lambda} \Delta \mathrm{OPL}({\bf r},t)
\end{equation}
Using the phase variations for each beam defined above, we can write the phase difference encoded into the illumination intensity as:
\begin{eqnarray}
\Delta \Phi({\bf r},t) & = & 2 \pi \, \nu_c \, t + 2 \pi \, \frac{n_2}{n_1} \, M \, \kappa_1 \, t \, \Delta x + \frac{n_2}{n_1} \, M \, \Delta k \, y \nonumber \\
	& & +  k_2 \, z \left[\sqrt{1 - \left(\frac{\lambda}{n_1} \, M \, \kappa_1 \, t \right)^2 - \left(\frac{n_2}{n_1} \, \frac{M \, \Delta k}{k_2} \right)^2} - 1 \right] + \Delta \Phi_a(t) + \alpha_1(t)
\label{eq:full_phase_diff}
\end{eqnarray}
where $\Delta \Phi_a(t) \equiv \Phi_\mathrm{pupil}(2 \pi \, \kappa_1 \, t) - \Phi_\mathrm{pupil}(0)$ is the aberration phase, the pupil phase at time $t$ referenced to the on-axis pupil phase value.

The illumination system is designed to allow portions of the $j=0$ and $j=1$ beams to pass at the equator of the objective lens (vertically), thereby removing the shear between the diffracted and undiffracted beams in the vertical dimension.  As such, one can make the approximation that in the object region, $\Delta k \rightarrow 0$ in the $y$-direction only, and the phase difference no longer varies with respect to the vertical dimension:
\begin{equation}
\Delta \Phi(x,z,t) = 2 \pi \, \nu_c \, t + 2 \pi \, \frac{n_2}{n_1} \, M \, \kappa_1 \, t \, \Delta x  +  k_2 \, z \left[\sqrt{1 - \left(\frac{\lambda}{n_1} \, M \, \kappa_1 \, t \right)^2} - 1 \right] + \Delta \Phi_a(t) + \alpha_1(t) 
\label{eq:phase_difference}
\end{equation}

Since the carrier frequency $\nu_c$ only appears in the phase difference, we can split the illumination intensity into a DC term and an AC term, which we denote as $I_0(t)$ and $I_1({\bf r},t)$ respectively. The illumination intensity is then $I_\mathrm{ill}({\bf r},t) = I_0(t) + I_1({\bf r},t)$, where:
\begin{eqnarray}
I_0(t) & = & v_{0,2}({\bf r},t) \, u_{0,2}({\bf r},t) + v_{1,2}({\bf r},t) \, u_{1,2}({\bf r},t) \nonumber \\ 
	 & = & a_0^2 + a_1^2 \, w^2(t)
\end{eqnarray} 
and
\begin{eqnarray}
I_1({\bf r},t) & = & v_{0,2}({\bf r},t) \, u_{1,2}({\bf r},t) + v_{0,2}({\bf r},t) \, u_{1,2}({\bf r},t) \nonumber \\
	& = & 2 \, a_0 \, a_1 \,  w(t) \, \cos \left[ \Delta \Phi ({\bf r},t) \right]
\end{eqnarray}
Note that the temporal dependence in $I_0(t)$ is due only to the temporal window, which accounts for a loss in transmitted amplitude of the $j = 1$ beam with scan time.  

Although the mathematical formalism at this point corresponds to a sinusoidal amplitude grating, i.e., $N =1$, experimentally the modulation mask is binary.  We therefore select the amplitude coefficients to reflect the diffractive amplitudes from a square grating: $a_0 = 1/2$, and $a_1 = 1/\pi$. Thus the illumination intensities are: 
\begin{eqnarray}
I_0(t) & = & \frac{1}{4} + \frac{1}{\pi^2} \, w^2(t) \\
I_1({\bf r},t) & = & \frac{1}{\pi} \, w(t) \, \cos \left[\Delta \Phi({\bf r},t) \right]
\label{eq:ac_intensity}
\end{eqnarray}

We can also express the measured signal as a sum of the AC and DC components:
\begin{equation}
S(t) = \left< \left[ I_0(t) + I_1({\bf r}, t) \right] \, C({\bf r}) \right >_{\bf r} = S_0(t) + S_1(t)
\end{equation}
The AC component of the illumination intensity is responsible for encoding the image into the temporal measurement from the single-element detector:
\begin{equation}
S_1(t) = \left< I_1({\bf r},t) \, C({\bf r}) \right>_{\bf r} = \frac{1}{\pi} \,  w(t) \left< \cos \left[\Delta \Phi({\bf r},t) \right] \, C({\bf r}) \right>_{\bf r}
\end{equation}


The illumination intensity pattern driven by the phase difference in \eqref{eq:full_phase_diff} varies with scan time in $x$ and $z$, but is constant with respect to the vertical coordinate, $y$.  This is expected, as the vertical spatial frequency on the modulation mask is constant with respect to time [\eqref{vertical_spatial_freq}] --  a result of being invariant with mask rotation angle, $\varphi_0(t)$. The effect of the vertical spatial frequency is to impart a linear phase delay in the vertical dimension that is directly proportional to the mask density, $\Delta k$, and another linear phase delay that arises from the shear of the re-imaged diffracted order with respect to the optic axis. The linear phase delay with respect to the axial coordinate leads to an intensity modulation in the case where the full $y$ spatial frequency extent of the zero-order and first diffracted order beam are allowed to pass through the CHIRPT illumination optical system. Placement of the filter that selects a portion of the $y$ spatial frequency content eliminates this modulation that arises from the vertical spatial frequency shear (Fig.~1c). The presence of the restrictive spatial filter leads to the illumination intensity distribution given in \eqref{eq:ac_intensity} with a phase difference defined by \eqref{eq:phase_difference}.

Finally, since only the lateral and axial components of the contrast function are encoded into the temporal signal, we assume that the contrast function is separable with respect to the vertical dimension, i.e., $C({\bf r}) = C_y(y) \, C(x,z)$, so that the AC signal is:
\begin{eqnarray}
S_1(t) & = & \frac{1}{\pi}\,  w(t)  \, \int_{-\infty}^{\infty} \! \!  \mathrm{d} y \, \, C_y(y)  \, \iint_{-\infty}^{\infty} \! \!  \mathrm{d} z  \, \mathrm{d} x  \, \, C(x,z) \nonumber \\ 
	& & \times \cos \! \left\{2 \pi \, \nu_c \, t +  2 \pi \, \left( \frac{n_2}{n_1} \, M \, \kappa_1 \, t \right) \, \Delta x  + k_2 \, z \, \left[ \sqrt{1 - \left( \frac{\lambda}{n_1} \, M \, \kappa_1 \, t \right)^2} -1 \right] + \Delta \Phi_a(t) + \alpha_1(t)  \right\}  
\label{spifi_signal_real}
\end{eqnarray}
Integration over the vertical coordinate in the expression above results in a constant multiplicative factor.  Let us define an overall prefactor to account for this integration and the leading factor of $1/\pi$ as $\gamma \equiv (1/\pi) \int \! \mathrm{d}y \, C_y(y)$.

The signal collected from the photodetector must be real-valued, and thus it follows that the signal displays conjugate spectral symmetry. Both the positive and negative fundamental modulation sidebands, corresponding to the AC component of $S(t)$, encode phase information, although they are conjugates to one another.  For image reconstruction, only one such signal is needed. The complex CHIRPT spatial frequency spectrum from the temporal data trace is isolated by filtering out the positive frequency sideband of the measured signal, then inverse Fourier transformed. 

After these operations, we arrive at an expression for the measured spatial frequency distribution; the amplitude of the the time-dependent trace contains the object intensity spatial frequency distribution, and the phase represents the propagation and modulation phase of the illumination intensity accumulated between the modulator and the object region.  In the analytic theory presented here, this is readily observed by representing the cosine term in \eqref{spifi_signal_real} as a sum of complex exponentials. To simplify the notation in the following expressions, we define the chirp parameter in the object region as: $\kappa_2 = (n_2/n_1) \, M \, \kappa_1$, and observe that the lateral spatial frequency in the object region is $f_{x,2}(t) = \kappa_2 \, t = k_{x,2}(t) /(2 \pi)$. Then we can write:
\begin{eqnarray}
S_1(t) & = & \frac{1}{2}\, w(t) \, \gamma \,   \iint_{-\infty}^{\infty} \! \!  \mathrm{d} z \, \mathrm{d} x  \, C(x,z)   \left\{ \exp \! \left[\ii \, 2 \pi \, \nu_c \, t + \ii \, k_{x,2}(t) \, \Delta x \frac{}{} \right. \right. \nonumber \\
	& &   \left. \left. + \, \ii \,  k_2 \, z \,\left( \sqrt{1 - \left[ \frac{k_{x,2}(t)}{k_2} \right]^2} -1 \right) + \ii \, \Delta \Phi_a(t) + \ii \,\alpha_1(t) \right]  +  \, \mathrm{c.c.} \right\}  
\end{eqnarray}
where c.c. denotes the complex conjugate.  We represent the positive and negative temporal sidebands centered at $\pm \nu_c$ as $S_{1+}(t)$ and $S_{1-}(t)$ respectively, such that $S_1(t) = S_{1+}(t) + S_{1-}(t)$. Each term results from a single sideband in the modulation frequency domain, and is therefore a complex quantity. 

While the intensity is by definition real-valued, each sideband can be thought of as resulting from the individual complex components of the intensity.  Denoting the complex components of the illumination intensity in the same manner as the measured signal, i.e., $I_1({\bf r},t) = I_{1+}({\bf r},t) + I_{1-}({\bf r},t)$, where $I_1({\bf r},t) \in \mathbb{R}$ and $I_{1\pm}({\bf r},t) \in \mathbb{C}$, we can write:
\begin{equation}
S_{1\pm}(t) = \left< I_{1\pm}({\bf r},t) \, C({\bf r}) \right>_{\bf r} = \left< v_{1,2}({\bf r},t) \, u_{0,2}({\bf r},t)  \, C({\bf r}) \right>_{\bf r} 
\end{equation} 
Notice that in order for this statement to hold, the complex components of the intensity must be conjugate to one another, i.e., $I_{1+}({\bf r},t) = \left[I_{1-}({\bf r},t) \right]^*$, and consequently the positive and negative temporal sidebands carry redundant spatial phase information. In practice, we select the positive temporal first harmonic for image reconstruction. Finally, we arrive at an expression for the image from a single lateral-axial plane encoded by CHIRPT:
\begin{eqnarray}
S_{1+}(t) & = & \frac{1}{2} \, w(t) \,  \ee^{\ii \, 2 \pi \, \nu_c \, t} \, \ee^{\ii \, \Delta \Phi_a(t)} \, \ee^{\ii \, \alpha_1(t)} \nonumber \\  
	& & \times \, \iint_{-\infty}^{\infty} \! \!  \mathrm{d} z \, \mathrm{d} x \, \, C(x,z) \, \exp \! \left\{ \mathrm{i} \, k_{x,2}(t) \, \Delta x \, + \ii \, k_2 \, z  \, \left[ \sqrt{1 - \left[ \frac{k_{x,2}(t)}{k_2} \right]^2} - 1 \right] \right\}
\label{eq:S1p}
\end{eqnarray}

\subsection{Spatial frequency projections}
CHIRPT encodes lateral image information by sequentially projecting an illumination pattern with a single lateral spatial frequency, $k_{x,2}(t)$, onto the specimen. The contrast function, $C(x,z)$ is measured by collecting the total radiant flux emanating from the specimen region for each spatial frequency.  The CHIRPT signal in time, $S_{1+}(t)$, represents a projection of the spatial frequency of the illumination intensity onto the object. Thus the CHIRPT signal can be recast into the lateral spatial frequency domain by relating it to the scan time. This is clear from \eqref{eq:S1p}. Rewriting the expression for the CHIRPT signal in lateral spatial frequency coordinates, we find:

\begin{eqnarray}
S_{1+}(f_x) & = & \frac{1}{2} \, w(f_x) \,  \ee^{\ii \, 2 \pi \, f_x \, x_c} \, \ee^{\ii \, \Delta \Phi_a(f_x)} \, \ee^{\ii \, \alpha_1(f_x)} \nonumber \\ 
	& & \times \int_{-\infty}^{\infty} \! \!  \mathrm{d} z \, \, \exp \left(\ii \, k_2 \,  z \,\left\{ \sqrt{1 - \left[ \frac{ f_{x,2}(t)}{f_2} \right]^2} - 1 \right\} \right) \nonumber \\
	& & \times \int_{-\infty}^{\infty} \! \!  \mathrm{d} x \, \,  C(x,z) \, \mathrm{e}^{\mathrm{i} \,2 \pi \,  f_x \, x} 
\label{temporal_signal}
\end{eqnarray}
where we have applied the shorthand notation $f_x = f_{x,2}(t)$ to represent the lateral spatial frequency of the collected signal. Notice that in this representation, the temporal window function limits the spatial frequency content of the measured signal, which ultimately sets a limit on the numerical aperture of the measurement.

The integration over the lateral coordinate has the form of a spatial Fourier transform of the contrast function with respect to the lateral coordinate, so the preceding expression can be written:
\begin{eqnarray}
S_{1+}(f_x) & = &  \frac{1}{2} \, w(f_x)  \,  \ee^{\ii \, 2 \pi \, f_x \, x_c} \, \ee^{\ii \, \Delta \Phi_a(f_x)} \, \ee^{\ii \, \alpha_1(f_x)} \nonumber \\
 & & \times \,  \int_{-\infty}^{\infty} \! \!  \mathrm{d} z \, \, \exp \left(\ii \, k_2 \,  z \,\left\{ \sqrt{1 - \left[ \frac{ f_{x,2}(t)}{f_2} \right]^2} - 1 \right\} \right) \, \tilde{C_x}(f_x,z)
\label{fourier_picture}
\end{eqnarray}
where $\tilde{C_x}(f_x,z)$ is the one-dimensional Fourier transform of $C(x,z)$ with respect to $x$.

Equations \eqref{eq:S1p}  and \eqref{fourier_picture} are descriptive of CHIRPT imaging, in that they demonstrate how amplitude and phase information from the spatial frequency domain are encoded into a temporal measurement.  By virtue of the phase delay with respect to the carrier frequency in the CHIRPT signal, the total spatial phase accumulated between the modulation plane and the object plane is encoded into a single temporal measurement.  This includes phase variation from defocus, which is explicitly included in the theory presented above, as well as higher-order spatial frequency phase variations.

Physically, one can think of the complex CHIRPT signal $S_{1+}(t) \leftrightarrow S_{1+}(f_x)$ as encoding the amplitude and phase of the lateral spatial frequency content, $f_x$.  A portion of the longitudinal spatial frequency content of the specimen is also encoded through the phase imparted by physical defocus. In other words, each temporal data point corresponds to the amplitude of a lateral spatial frequency $f_{x,2}(t)$, and a unique axial spatial frequency $f_{z,2}(t)$, because the axial spatial frequency of a plane wave is determined by the transverse spatial frequency and the total wavenumber constrained by the dispersion relationship dictated by the Helmholtz equation. This point is shown explicitly in Section~\ref{sec:transfer_function}.

\section{Point emitter signal}
It is instructive to study the case of a point emitter, which is formally equivalent to examining the transfer function in CHIRPT imaging.  The contrast function for a point emitter located at $(x_p,z_p)$ in the $(x,z)$-plane, relative to the centroid position $(x_c,z_0)$, is described by a product of Dirac-$\delta$ functions: $C(x,z) = \delta(x-x_p) \delta(z-z_p)$.  Note that for simplicity we have omitted proportionality constants such as the absorption cross section.  Assuming no pupil phase for the imaging system ($\Delta \Phi_a(t) = 0$) and zero translation of the mask with scan time ($\alpha_1(t)$), \eqref{eq:S1p} reduces to:
\begin{equation}
S_{1+}(t) = \frac{1}{2} \, w(t) \, \exp \left[ \mathrm{i} \, 2 \pi \,\left( \nu_c +  \kappa_2 \, x_p \right) \, t \right] \, \exp \left[ \mathrm{i} \, k_2 \, z_p \, \left\{ \sqrt{1 - \left[\frac{k_{x,2}(t)}{k_2} \right]^2} - 1 \right\} \right]
\label{point-emitter}
\end{equation}
It is clear that the temporal first harmonic of the photodetector signal encodes lateral position of a single emitter through carrier frequency, and axial position through frequency chirp. In the paraxial limit, where $\left[k_{x,2}(t)/k_2 \right] \ll 1$, we can make the approximation:
\begin{equation}
\sqrt{1 - \left[\frac{k_{x,2}(t)}{k_2} \right]^2} \approx 1 - \frac{1}{2} \left[\frac{k_{x,2}(t)}{k_2} \right]^2 = 1 - \frac{1}{2} \left( \frac{\lambda}{n_2} \, \kappa_2 \, t \right)^2
\end{equation}
and \eqref{point-emitter} becomes:
\begin{equation}
S_{1+}(t) \approx \frac{1}{2} \, w(t) \, \exp \left[ \mathrm{i} \, 2 \pi \,\left( \nu_c +  \kappa_2 \, x_p \right) \, t \right] \, \exp \left[ -\ii \, \pi \, \frac{\lambda}{n_2} \, \left( \kappa_2 \, t \right)^2  \, z_p\right]
\end{equation}
Note that in the paraxial approximation, axial phase manifests in the form of a Fresnel zone plate (FZP) in the lateral spatial frequency domain, suggesting a strong analogy to off axis spatial holography \cite{Leith:62,Leith:1963wn}.

For a collection of $N_p$ point emitters with locations given by a set of coordinates $\left\{ (x_p,z_p) \right\}$, the contrast function becomes:
\begin{equation}
C(x,z) = \sum_{p=1}^{N_p} \, a_p \, \delta \left(x - x_p \right) \, \delta \left(z - z_p \right)
\end{equation}
and the signal from the photodetector is a linear combination of the signals from each emitter. In the paraxial approximation, the temporal scan encodes a summation of FZPs in the lateral spatial frequency domain because the total signal is a linear combination of the signal from each emitter \cite{Rogers:1950jk}.  Uniqueness of the illumination pattern for conjugate locations in the object plane, e.g., $(x_1,z_1)$ and $(x_1,-z_1)$, is maintained by the shift in instantaneous frequency relative to the carrier frequency, $\nu_c$, preventing ambiguity of emitter location in the measured signal.

\subsection{Instantaneous modulation frequency}
The temporal modulation frequency of the positive first harmonic signal is dependent upon the position of the point emitter in the lateral-axial plane.  This can be made explicit by computing the form of the instantaneous modulation frequency, $\nu(t)$:
\begin{eqnarray}
\nu(t) & = &  \frac{\partial}{\partial \, t} \left\{ \frac{\Delta \Phi(x_p,z_p,t)}{2 \pi}  \right\} = \frac{\partial}{\partial \, t} \left\{ \frac{\Delta \mathrm{OPL}(x_p,z_p,t)}{\lambda}  \right\}  \nonumber \\
	& = & \nu_c +  \kappa_2 \, x_p - \frac{\lambda}{n_2} \, \frac{\kappa_2^2 \, t}{\sqrt{1 - \left(\frac{\lambda}{n_2} \, \kappa_2 \, t \right)^2}} \, z_p   + \, \frac{1}{2 \pi} \, \frac{\partial \, \Delta \Phi_a(t)}{\partial \, t}  + \frac{1}{2 \pi} \, \frac{\partial \, \alpha_1(t)}{\partial \, t}  
\end{eqnarray}
This relationship between the position of the point emitter and the temporal modulation frequency of the measured signal is convenient for understanding how each position has a unique temporal modulation pattern in the object plane.

\begin{figure}[ht]
\begin{center}
\resizebox{3.5in}{!}{\includegraphics{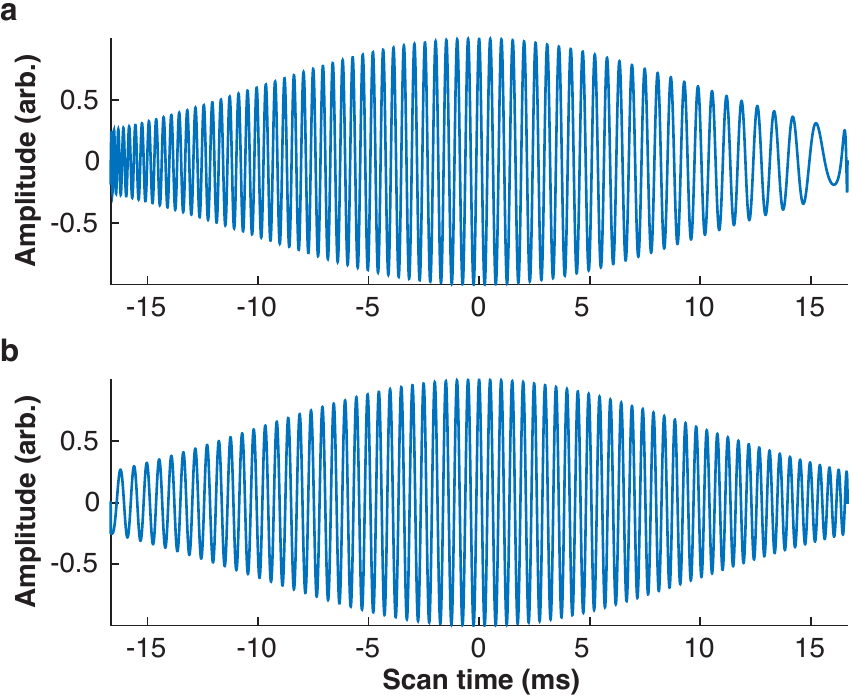}}
\caption{\label{time_traces} The instantaneous modulation frequency varies with the position of each emitter in the object region. Here we plot the real portion of the complex CHIRPT signal computed for (a) $z_p = 4.9~\mu \mathrm{m}$ and (b) $z_p = -1~\mu \mathrm{m}$. Parameters of the imaging system simulated here can be found in the main text. }
\end{center}
\end{figure}

Figure~\ref{time_traces} shows the real portion of the complex temporal CHIRPT traces, $\mathrm{Re}\left\{S_{1+}(t) \right\}$, computed for a single emitter with two different defocus values.  For simplicity, we set the lateral position of the emitter to zero ($x_p = 0$), assumed a perfect image relay system for which the aberration phase is zero, and set the phase due to lateral translations of the disk to zero. The signal was computed for a CHIRPT microscope with $\lambda = 532~\mathrm{nm}$, $M = 95$, $\Delta k = 70/\mathrm{mm}$, and an air immersion illumination objective with NA = 0.8. While the typical carrier frequency for such a system would be on the order of 10-100~kHz (depending on the rotation frequency of the modulator mask), we set the carrier frequency to 2~kHz to make it possible to visualize the chirp in modulation frequency as a function of scan time. A Gaussian temporal window function with full width at $1/e$ point of 20~ms was also included. 

The real component of the complex temporal signals in Fig.~\ref{time_traces} show how a one-dimensional temporal measurement encodes the emitter location in two dimensions by deviations from the carrier frequency.  In the analytic example shown here, the magnitude and sign of frequency chirp determine the emitter location $(x_p,z_p)$.  In general, other phase disturbances, such as systematic aberration phase, are also encoded into the modulation frequency chirp. In practice a single emitter can be used to empirically determine the systematic aberration phase of the microscope, allowing for digital aberration correction to be applied to images collected with CHRIPT without the need for iterative algorithms \cite{Field:2015fr}.

\begin{figure}[ht]
\begin{center}
\resizebox{3.5in}{!}{\includegraphics{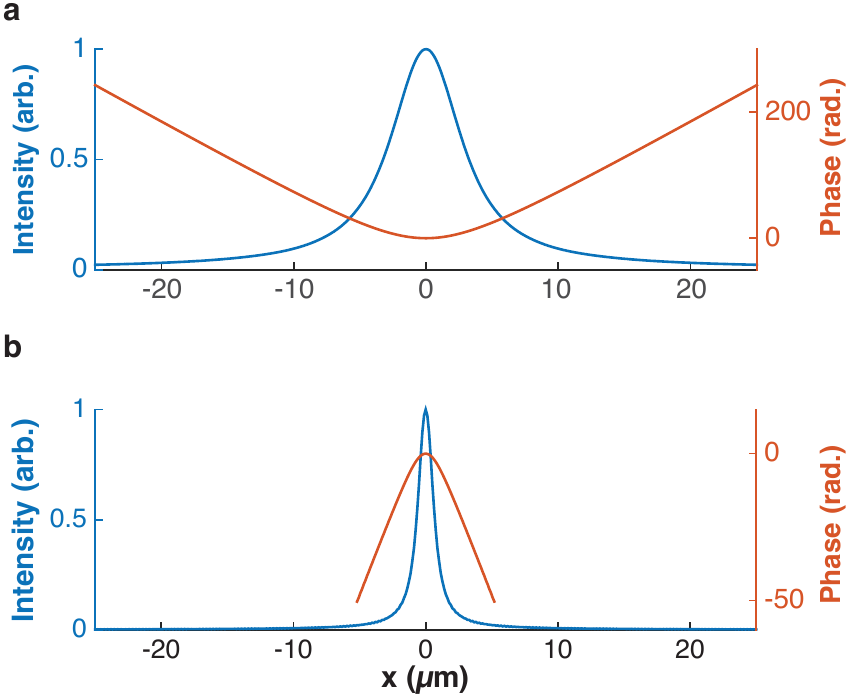}}
\caption{\label{1d_images} One-dimensional images computed from the complex CHIRPT data in Fig.~\ref{time_traces}. Both the intensity of the contrast signal and the phase difference between the two illumination beams are encoded for each image. (a) The image computed with $z_p = 4.9~\mu \mathrm{m}$ shows positive concavity in the spatial phase, consistent with the positive chirp observed in Fig.~\ref{time_traces}a. (b) Similarly, the image computed for $z_p = -1~\mu \mathrm{m}$ shows negative concavity in the spatial phase. The width of the distribution is much more narrow in this image because the magnitude of the defocus is significantly smaller. }
\end{center}
\end{figure}

Each temporal CHIRPT trace in time represents a one-dimensional line image in the lateral dimension. The image is recovered by demodulating the temporal traces by the average frequency, converting the scan time to lateral spatial frequency via $f_{x,2}(t) = (n_2/n_1) \, M \, \kappa_1 \, t$, and computing the Fourier transform with respect to $f_{x,2}$. Line images corresponding to temporal traces displayed in Fig.~\ref{time_traces} are shown in Fig.~\ref{1d_images}, where a fast Fourier transform (FFT) was used to compute the images from the numeric temporal data. 

\section{CHIRPT transfer function \label{sec:transfer_function}}
The expression for the one-dimensional CHIRPT signal for a point emitter can be used to compute the two-dimensional transfer function in the spatial frequency domain, $(f_x,f_z)$. As shown in \eqref{temporal_signal}, the lateral spatial frequency, $f_x$, can be substituted in place of the scan time, $t$. Making this substitution, we express \eqref{point-emitter} as:
\begin{equation}
S_{1+}(f_x)  =  \frac{1}{2} \, w(f_x) \, \exp \! \left\{ \mathrm{i} \, 2 \pi \, f_x \, x_p  \frac{}{}^{\,}  + \, k_2 \, z_p  \left[ \sqrt{1 - \left(\frac{\lambda}{n_2} \, f_x \right)^2} - 1 \right] \right\}
\end{equation}
where we have demodulated the signal by the carrier frequency for simplicity. The transfer function for CHIRPT imaging under single-frequency illumination can be computed with a spatial Fourier transform of this expression with respect to $z$.  Since the lateral position of the emitter contributes only a linear phase ramp, we  set $x_p = 0$ for simplicity; then the transfer function is:
\begin{eqnarray}
\mathcal{H}(f_x,f_z) & = & \frac{1}{2} \,  w(f_x) \, \int_{-\infty}^{\infty} \! \! \mathrm{d}z \, \, \mathrm{e}^{\mathrm{i} \, k_2 \, z \, \left[\sqrt{1 - \left(\frac{\lambda}{n_2} \, f_x \right)^2} - 1\right] } \, \mathrm{e}^{-\mathrm{i} \, 2 \pi \, f_z \, z}  \nonumber \\ 
	& = & \frac{1}{2} \, w(f_x) \, \int_{-\infty}^{\infty} \! \! \mathrm{d}z \, \, \exp \! \left[\mathrm{i} \, 2 \pi \, \left(f_z - \Delta f_z \right) \, z \right] \nonumber \\
	& \propto & w(f_x) \,  \delta \left(f_z - \Delta f_z \right)
\label{transfer_function}
\end{eqnarray}
where $\Delta f_z$ represents the difference in longitudinal spatial frequency between the diffracted and undiffracted beams, and $\delta(\cdot)$ represents the Dirac-delta function. The Dirac-$\delta$ function is non-zero only when its argument is equal to zero, so the transfer function is zero everywhere in the spatial frequency domain except for where $f_z = \Delta f_z$. Thus we can write the longitudinal spatial frequency in the object domain, which is constrained by the dispersion relationship imposed by the Helmholtz equation for plane wave propagation:
\begin{equation}
f_{z,2}(t) = \frac{n_2}{\lambda}\left\{\sqrt{1 - \left[ \frac{\lambda}{n_2} f_{x,2}(t) \right]^2} - 1 \right\}
\label{fz_total}
\end{equation} 
Since there is a unique mapping of the lateral and axial spatial frequency information, each temporal measurement $S_{1+}(t)$ defines a two-dimensional plane wave. Summation of all the plane waves over the duration of a temporal scan forms a 2D image from a 1D data trace.

Equation~\eqref{transfer_function} elucidates a fundamental limitation of CHIRPT imaging, and coherent diffractive imaging in general -- the transfer function contains vanishingly small longitudinal frequency support, and thereby the imaging system does not display axial sectioning.  In his seminal paper from 1969, E.~Wolf notes that holography captures a portion of the complex object response that is determined by the illumination conditions \cite{Wolf:1969tea}. Since that information is confined to the surface of the Ewald sphere in the spatial frequency domain, the possibility of three-dimensional image reconstruction with a single view is prohibited. The radius of the Ewald sphere is defined by the wavenumber of the light.  Noting that the wavenumber can be related to the linear spatial frequency by $k_2 = 2 \pi \, f_2 = 2 \pi \, (n_2/\lambda)$, \eqref{fz_total} can be rearranged to show:
\begin{equation}
f_2^2 = f_{x,2}^2 + \left(f_{z,2} + f_2\right)^2
\end{equation}
which represents a slice through the Ewald sphere at $f_y = 0$ and shifted in longitudinal frequency by $f_2$. The Ewald circle described by this expression is shifted by $f_2$ in the axial dimension because CHIRPT encodes the difference between the diffracted and undiffracted beams. Equation~\eqref{k_ewald} represents the full Ewald sphere for an electric field diffracted from the mask. 

\section{Experimental CHIRPT data}
To validate the theoretical analysis presented here, we imaged a 100~nm-diameter fluorescent nanodiamond (FND; ND-400NV-100nm-10mL, Ad\'{a}mas Nanotechnologies) as a function of defocus. Since the FND was sub-diffraction limited for the CHIRPT microscope, it acted as a point emitter and allowed for the transfer function of the microscope to be determined. A laser with 532~nm nominal wavelength (Sprout 5G, Lighthouse Photonics) was brought to a horizontal line focus with a 150~mm focal length cylindrical lens (ACY254-150-A-ML, ThorLabs) onto the spinning modulator.  A modulation pattern with $\Delta k = 70/\mathrm{mm}$ was printed in aluminum onto a glass disk (Projection Technologies). The modulation plane was image relayed to the specimen region in two stages to allow unimpeded access to a conjugate of the objective lens pupil plane for filtering with the horizontal slit.  The image relay system consisted of 125 mm and a 100 mm focal length lenses (ACH-254-125-ML and ACH-254-100-ML, ThorLabs) for the first stage, and a 250 mm focal length lens (ACH-254-250-ML, ThorLabs) and 0.8~NA objective lens (Zeiss, N-Achroplan 50x/0.8 NA Pol) for the second stage.  The overall system magnification was 95$\times$. Fluorescence was collected in the epi direction by reflecting the illumination beams into the specimen with a dichroic beamsplitter (FF562-Di03-25x36, Semrock) and collecting fluorescence transmitted through the dichroic.  Fluorescence emission was filtered with a bandpass filter (ET645/75m, Chroma). Fluorescence intensity was measured with a photomultiplier tube (PMT; H7422P-40, Hamamatsu).  The signal was amplified with a low-noise current amplifier (DHPCA-100, Femto) and digitized with a data acquisition card (PCI-6110, National Instruments). Specimen positioning was controlled with a three-dimensional motorized stage ((x,y): MS-2000; z: LS-50, Applied Scientific Instruments).  Data collection and stage control were implemented in a custom C\# application written in house. 

\begin{figure}[!htb]
\begin{center}
\resizebox{3.5in}{!}{\includegraphics{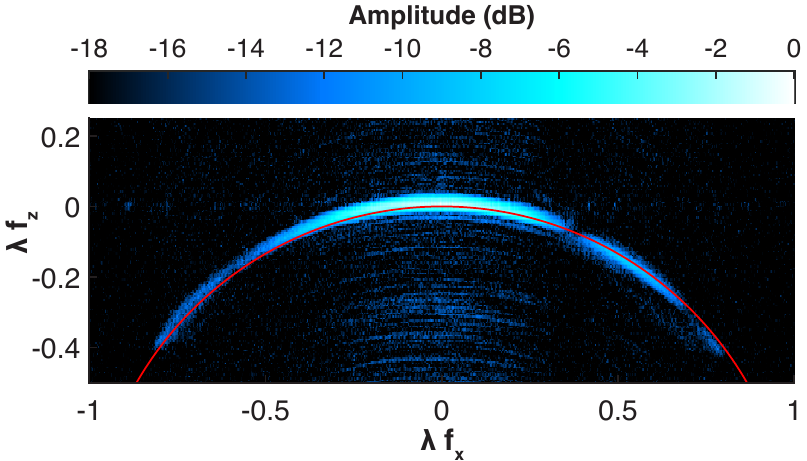}}
\caption{\label{ewald_measured} Experimental transfer function from a CHIRPT microscope, measured with a sub-diffraction limited FND. The theoretical Ewald phase is plotted in red. }
\end{center}
\end{figure}

Figure~\ref{ewald_measured} shows the average OTF computed from 20 images of the FND as a function of defocus, $\Delta z$. The overlaid red line represents the Ewald phase, $\Delta \phi_{\mathrm{Ewald}}(f_x,f_z)$, computed with the experimental parameters described above. The excellent agreement between the measured and computed transfer function verifies the theoretical analysis.

\section{Emitters of finite extent}
So far we have computed the CHIRPT signal for point emitters.  However this is only valid in the case where a physical emitter is smaller than the diffraction-limited resolution of a CHIRPT microscope.  To explore CHIRPT imaging in the case of finite-extent objects, we consider a contrast function defined by a two-dimensional Gaussian distribution in the $(x,z)$ plane:
\begin{equation}
C(x,z) = \beta \, \exp \left\{- \left[ \frac{(x - x_p)^2}{w_x^2} + \frac{(z - z_p)^2}{w_z^2} \right]  \right\}
\end{equation}
Note that factors such as the absorption cross section are absorbed into the scalar constant, $\beta$. Inserting this contrast function in \eqref{eq:S1p}, we find:
\begin{eqnarray}
S_{1+}(t) & = & \frac{1}{2} \, w(t) \, \pi \, \beta \, w_x \, w_z \, \mathrm{e}^{- w_x^2 \, \left(\pi \, \kappa_2 \, t \right)^2} \, \mathrm{e}^{w_z^2 \, \left(\pi \, \kappa_2 \, t  \right)^2 } \, \mathrm{e}^{\frac{1}{2}\, \left( k_2 \, w_z\right)^2 \left[\sqrt{1 - \left( k_{x,2}(t) / k_2 \right)^2} - 1 \right]} \nonumber \\
	&  & \times \,  \mathrm{e}^{\mathrm{i} \, 2 \pi \, \nu_c \, t} \, \mathrm{e}^{\mathrm{i} \, \Delta \Phi_a(t)} \,\mathrm{e}^{\mathrm{i} \, 2 \pi \, \left( \kappa_2 \, t \right) \, x_p} \, \mathrm{e}^{\mathrm{i} \, k_2 \, z_p \left[\sqrt{1 - \left(k_{x,2}(t) / k_2 \right)^2} - 1 \right]}
\label{gaussian-emitter}
\end{eqnarray}
The complex component of the temporal CHIRPT signal in \eqref{gaussian-emitter} is identical to that of the signal for a point emitter we have previously derived [\eqref{point-emitter}].  In addition to this complex term, the amplitude of the temporal signal is modulated with a real function that depends upon the size of the emitter in both the lateral and axial dimensions, expressed through the Gaussian widths $w_x$ and $w_z$ respectively. As such we can write:
\begin{equation}
S_{1+}^{\mathrm{(Gaussian)}}(t) = \pi \, \beta \, w_x \, w_z \, A(t;w_x,w_z) \, S_{1+}^{\mathrm{(}\delta\mathrm{)}}(t)
\end{equation}
where $S_{1+}^{\mathrm{(}\delta\mathrm{)}}(t)$ is equivalent to the expression in \eqref{point-emitter}, and the amplitude of the complex temporal signal depends on the parameters of the Gaussian emitter:
\begin{equation}
A(t;w_x,w_z) = \exp \left[ - \left(\pi \, \kappa_2 \, t \right)^2 \, \left(w_x^2 - w_z^2 \right) \right]  \exp \left\{ \frac{1}{2}\, \left( k_2 \, w_z\right)^2 \left[\sqrt{1 - \left[ \frac{k_{x,2}(t)}{ k_2} \right]^2} - 1 \right] \right\} 
\end{equation}
To understand how this amplitude modulation affects the measured image, it is useful to covert the notation to the lateral spatial frequency domain:
\begin{equation}
A(f_x;w_x,w_z) = \exp \left[ - \left(\pi \, f_x \right)^2 \, \left(w_x^2 - w_z^2 \right) \right] \exp \left  \{ \frac{1}{2}\, \left( k_2 \, w_z\right)^2 \left[\sqrt{1 - \left( \frac{\lambda}{n_2} \,f_x \right)^2} - 1 \right] \right\}
\end{equation}
In this form, the amplitude modulation describes the finite support of the measured CHIRPT image in the lateral spatial frequency domain. Several properties of this function are worth noting.  

First, there is a Gaussian dependence on the inverse of the width of the emitter in the lateral dimension.  This is fully expected as the CHIRPT trace in time physically corresponds to a measurement of the lateral spatial frequency content of the object.  As the width of the object increases, fewer lateral spatial frequencies are needed to reproduce an image of the object, and consequently the finite support is reduced.

There is a similar dependence on the axial width of the object, although there is additional term that modifies the finite support which follows an Ewald-like form.  The axial component of the finite support has a profound affect on images collected with CHIRPT, as it implies that the axial size of an object limits the lateral resolution of the image. For example, consider an object for which $w_x \rightarrow 0$, but $w_z$ is larger than the diffraction limit. The frequency support becomes:
\begin{equation}
\lim_{w_x \rightarrow 0} A(f_x;w_x,w_z) = \mathrm{e}^{-\pi^2 \, w_z^2 \, f_x^2} \, \mathrm{e}^{\frac{1}{2}\, \left( k_2 \, w_z\right)^2 \left[\sqrt{1 - \left( \frac{\lambda}{n_2} \,f_x \right)^2} - 1 \right]}
\end{equation}
Clearly the finite support is reduced by the amplitude modulations due to the finite width of the object in the axial dimension.

Finally, we note that in the limit that both $w_x \rightarrow 0$ and $w_z \rightarrow 0$, the amplitude function, tends towards unity, and the temporal trace, when normalized by the object widths, reproduces the form computed for a point emitter previously.
\begin{equation}
\lim_{\left\{w_x,w_z\right\} \rightarrow 0} S_{1+}^{\mathrm{(Gaussian)}}(t) = S_{1+}^{\mathrm{(}\delta\mathrm{)}}(t)
\end{equation}

\section{Comparison with SPIFI}
We have shown that in order to achieve a unique mapping of axial location with lateral spatial phase only one diffracted beam is passed through the imaging system and allowed to interfere with the undiffracted beam in the object region.  If both of the diffracted beams ($j = \pm 1$) are allowed to pass the imaging system, the CHIRPT microscope becomes equivalent to a SPIFI microscope \cite{Schlup:2011gpa,Futia:2011tra,Higley:2012ce,Howard:2012gca,Higley:2013vfa}. Due to the conjugate symmetry of the axial phase component for the diffracted orders, the transfer function for SPIFI is symmetric about the lateral spatial frequency axis.  Consequently, the illumination pattern is not unique for all points in the $(x,z)$ plane, and the axial position of an emitter cannot be uniquely determined.

The temporal first harmonic of the photodetector signal for a point emitter measured with SPIFI can be expressed by considering the field combinations that contribute to the sideband centered at $+\nu_c$. The corresponding complex portion of the illumination intensity is $I_{1+}({\bf r},t) = v_{1,2}({\bf r},t) \, u_{0,2}({\bf r},t) + u_{-1,2}({\bf r},t) \, v_{0,2}({\bf r},t)$, and the complex SPIFI signal at the positive sideband is:
\begin{eqnarray}
S_{1+}(t) & = & \left< I_{1+}({\bf r},t) \, C({\bf r}) \right>_{\bf r} = \left< v_{1,2}({\bf r},t) \, u_{0,2}({\bf r},t) \, C({\bf r}) \right>_{\bf r} + \left< u_{-1,2}({\bf r},t) \, v_{0,2}({\bf r},t) \, C({\bf r})\right>_{\bf r} \nonumber \\
	& = & a_0 \, a_1 \, w(t) \left< \exp \left[ \ii \left(\Phi_{1,2}({\bf r},t) - \Phi_{0,2}({\bf r},t) \right) \right] \, C({\bf r}) \right>_{\bf r} \nonumber \\
	& & + a_0 \, a_1 \, w(t) \left< \exp \left[ \ii \left(-\Phi_{-1,2}({\bf r},t) + \Phi_{0,2}({\bf r},t) \right) \right] \, C({\bf r}) \right>_{\bf r}
\end{eqnarray}
Following the formalism derived in Section \ref{sec:chirpt_imaging}, it is straightforward to show that the phase differences in the expression above have conjugate symmetry in both lateral spatial frequency and axial spatial frequency. In other words, instead of a single measurement in time corresponding to a unique lateral and axial frequency pair, $(f_x,f_z)$, the spatial frequency pair $(f_x,-f_z)$ is simultaneously encoded. Explicitly, the SPIFI signal for a point emitter is:
\begin{eqnarray}
S_{1+}(t) & = & \frac{1}{2} \, w(t) \, \exp \left[ \mathrm{i} \, 2 \pi \,\left( \nu_c + \kappa_2 \, x_p \right) \, t  \right] \exp \left[\mathrm{i} \, k_2 \, z_p \, \left\{ \sqrt{1 - \left[\frac{k_{x,2}(t)}{k_2} \right]^2} - 1 \right\} \right]  \nonumber \\
	& &  + \frac{1}{2} \, w(t) \, \exp \left[ \mathrm{i} \, 2 \pi \,\left( \nu_c + \kappa_2 \, x_p \right) \, t  \right] \exp \left[-\mathrm{i} \, k_2 \, z_p \, \left\{ \sqrt{1 - \left[\frac{k_{x,2}(t)}{k_2} \right]^2} - 1 \right\} \right] \nonumber \\
	& = & w(t) \, \exp \left[ \mathrm{i} \, 2 \pi \,\left( \nu_c + \kappa_2 \, x_p \right) \, t  \right]  \, \cos \left[ k_2 \, z_p \, \left\{ \sqrt{1 - \left[\frac{k_{x,2}(t)}{k_2} \right]^2} - 1 \right\}  \right]
\end{eqnarray}
From this expression it is clear that the uniqueness of the axial phase is destroyed if both diffracted orders are allowed to contribute to the illumination intensity pattern. As a result, a point emitter imaged with SPIFI results in a temporal pattern that is similar to the well known twin-image problem in holography, wherein an object is known to be defocused but the sign of the defocus is unknown. 

\begin{figure}[!ht]
\begin{center}
\resizebox{5in}{!}{\includegraphics{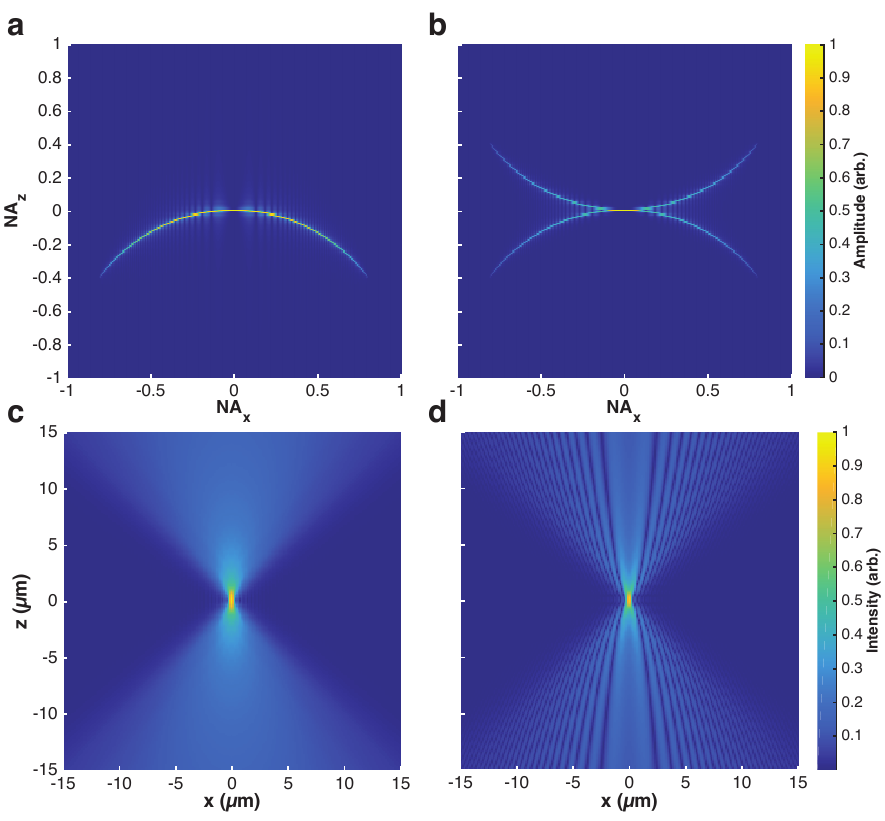}}
\caption{\label{psf_otf_comparison} Theoretical OTF and PSF for CHIRPT and SPIFI.  (a) The CHIRPT OTF follows an arc in the spatial frequency domain defined by the differential axial phase, $\Delta \phi_{\mathrm{Ewald}}(f_x,f_z)$. Here the numerical aperture in each dimension is defined by $\mathrm{NA}_{\{x,z\}} = \lambda \, f_{\{x,z\}}$. (b) Conversely, the SPIFI OTF contains two arcs with conjugate symmetry in longitudinal spatial frequency. (c) The CHIRPT PSF contains unique spatial phase information in the $(x,z)$ plane. (d) The SPIFI PSF contains modulations in the intensity pattern which result from the degeneracy in axial spatial frequency content.}
\end{center}
\end{figure}

We computed the transfer function and PSF for CHIRPT and SPIFI imaging (Fig.~\ref{psf_otf_comparison}). While the CHIRPT transfer function displays a unique mapping of each point in the spatial frequency domain, the SPIFI data shows that for each lateral spatial frequency there are two possible axial spatial frequencies of equal magnitude but opposite signs. This is a result of the cosine in the axial dimension. Similarly, the PSFs for CHIRPT and SPIFI are distinctly different -- whereas CHIRPT has a smooth PSF about the axial dimension, the SPIFI PSF displays spatial modulations.

\section{Discussion}
We have presented a theoretical analysis of CHRIPT imaging, a new form of optical microscopy that captures the spatial phase of a specimen with a measurement of incoherent (e.g., fluorescent) light. The analysis presented here assumes plane wave illumination with a single-frequency, spatially coherent source.  The analytic expressions developed here are valid for CHIRPT microscope systems in which focusing in the vertical dimension can be ignored.  We have verified the analytic expressions with experimental data.

CHIRPT makes it possible to directly apply the powerful suite of coherent diffraction imaging (CDI) tools -- which rely on spatial phase information available with coherent light -- to images formed with incoherent light for the first time. CHIRPT opens a window to deep, large volume, and high speed {\em in vivo} imaging that will bring new scientific insights to a range of fields. These capabilities are poised to dramatically impact biological imaging in particular.

\section*{Funding Information}
Funding was provided by the W.M.~Keck Foundation.

\section*{Acknowledgments}
We thank Susanta Sarkar from Colorado School of Mines for generous donation of the FND samples used in this work.

\bibliography{refs}


\end{document}